\documentclass[showpacs,aps,pre,twocolumn,superscriptaddress,textsuperscript]{revtex4}
\usepackage{graphicx}
\usepackage{amssymb}
\usepackage[sort&compress]{natbib}
\begin{document}
\title {Semimetallic single atomic layer silicon allotrope with Dirac fermions}
\author{Haiping Wu}
\email[Corresponding author:]{mrhpwu@njust.edu.cn}
\affiliation{Department of Applied Physics, Nanjing University of Science and Technology, Nanjing 210094, China}
\author{Yan Qian}
\email[Corresponding author:]{qianyan@njust.edu.cn}
\affiliation{Department of Applied Physics, Nanjing University of Science and Technology, Nanjing 210094, China}
\author{Zhengwei Du}
\affiliation{Department of Applied Physics, Nanjing University of Science and Technology, Nanjing 210094, China}
\author{Renzhu Zhu}
\affiliation{Department of Applied Physics, Nanjing University of Science and Technology, Nanjing 210094, China}
\author{Erjun Kan}
\affiliation{Department of Applied Physics, Nanjing University of Science and Technology, Nanjing 210094, China}
\author{Kaiming Deng}
\affiliation{Department of Applied Physics, Nanjing University of Science and Technology, Nanjing 210094, China}

\date{\today}
\begin{abstract}
Materials with Dirac point are so amazing since the charge carriers are massless and have an effective speed of light. Among the reported two-dimensional silicon allotropes, no one showing such exciting nature was proved experimentally. This fact motivates us to search for other such two-dimensional silicon allotropes. As a result, a new single atomic layer thin silicon allotrope was predicted by employing CALYPSO code in this work. This silicon allotrope is composed of eight-membered rings linked by Si-Si bonds and presents buckling formation. Expectedly, the electronic calculation reveals that there exists Dirac point at Fermi energy level. Furthermore, the ab initio molecular dynamics simulations displays that the original atomic configuration can be remained even at an extremely high temperature of 1000 \textit{K}. We hope this work can expand the family of single atomic layer thin silicon allotropes with Dirac fermions.
\end{abstract}
\maketitle
\section{Introduction}
Si-based materials are still most important for the modern industry. Simulated by the discovery of graphene\cite{Novoselov} and demand of miniaturization on electric components, a lot of work is pushed into searching for low-dimensional materials, and many such materials have been reported theoretically or experimentally\cite{Neto}. As a analogue of graphene, silicene was first theoretically predicted by Cahangirov in 2009\cite{Cahangirov}. The discovery of two-dimensional (2D) silicon shows so exciting, since it is the only one that can match well with the previous Si-based devices. Besides, the free-standing silicene also displays the special properties different from the corresponding bulk counterpart, for instance, there exists a linear energy dispersion near the Fermi energy level (\textit{E}$_{F}$), which indicates that the charge carriers can transport like massless Dirac fermions\cite{Cahangirov}.

However, silicene is unstable in the air due to the dangling bond, resulting in the fact that silicene is always generated on some suitable substrates in experiment, such as Ag(111), Ir(111), ZrB$_{2}$(0001), and Ru(0001) bulk substrates\cite{Chen,Lalmi,Feng,Vogt,Fleurence,Meng,Huang}. Unfortunately, no experimental observation can directly verify that there exists Dirac point in silicene deposited on substrates. We think the reason is that the interaction between the substrates and silicene heavily affect the property of silicene and even destroy the Dirac point. On the other hand, many theoretical works are focused on the free standing 2D silicon allotropes and many have been reported\cite{Gimbert,Cahangirov2,Matusalem,ZWang,Wu}, although the number is much less than that of the 2D carbon allotropes\cite{Ivanovskii}. Among these predicted allotropes beyond silicene, only the one given by Wu\cite{Wu} is single atomic layer thin, the others all are multi atomic layers thick. Notably, most of these materials presents metallic or semiconducting properties, very few of them show the character of Dirac point. Up to our knowledge, there are only two theoretically designed 2D silicon allotropes possessing Dirac point. One is named siliconeet predicted by Wang et al\cite{ZWang} and it is four atomic layers thick, and the other one is single atomic thin and generated on c-BN(111) substrate\cite{Wu}.

Although many 2D silicon allotropes have been theoretically reported, but no literature reported that any of these 2D silicon allotropes has been experimentally confirmed, except silicene which is single atomic layer thin. The reason maybe is that the atomic configuration with multi atomic layers is hard to be generated on substrates, and as a substrate, c-BN is too hard to make a clean c-BN(111) surface with previous technology. Additionally, the dynamical and thermal stabilities of some mentioned 2D silicon allotropes, as reported in Ref \cite{Matusalem}, are still an open question, while these properties would deeply affect the possibility of experimental synthesis.

The above facts encourage us to search for other stable 2D single atomic thin silicon allotropes with Dirac fermions, since single atomic thin 2D materials are relatively easy to be deposited on substrates. With the help of CALYPSO code, a stable buckling single atomic thin 2D silicon allotrope is synthesized in this work, and further calculation indicates that there presents Dirac point at \textit{E}$_{F}$ as expected. This discovery effectively expands the family of 2D silicon materials with Dirac fermions.

\section{Computational methods}
 Firstly, CALYPSO code is employed to search for 2D silicon allotropes. The code is very one that developed to search for the stable structures of compounds by using the swarm-intelligence based structural prediction calculations\cite{wang1,wang2}. The underlying ab initio structural relaxations and electronic band structure calculations are carried out in the framework of DFT within generalized$-$gradient approximations using the Perdew$-$Burke$-$Ernzerhof (PBE) exchange$-$correlation functional and projector augmented wave (PAW) potentials\cite{Perdew, Kresse1}, and these performances are all implemented in VASP code\cite{Kresse2}. The structural relaxations are performed until the Hellmann-Feynman force on each atom is less than 0.001 eV/{\AA}. To ensure high accuracy, the k-point density and the plane waves cutoff energy are increased until the change of the total energy is less than 10$^{-5}$ eV, and the Brillouin$-$zone (BZ) integration is carried out using 15$\times$15$\times$1 Monkhorst$-$Pack grid in the first BZ, the plane waves with the kinetic energy up to 600 eV is employed. In addition, the simulations are performed using a 2$\times$2$\times$1 supercell based on unit cell, and the repeated layered geometry is with a thick vacuum region of 20 {\AA}. The phonon calculations are performed using a supercell approach implemented in the PHONOPY code\cite{K,A}.

\section{Results and discussions}
By employing CALYPSO code, a number of 2D silicon allotropes were formed. Among the ones with low energies, a configuration was found to be one atomic layer thin and with buckling character (named silicoctene), and the structure is pictured in Fig. 1(a). Owing to the fact that Si prefers to adopt \textit{sp}$^{3}$ hybridization as in silicene, the buckling structure of silicoctene is quite reasonable. As drawn in Fig. 1(a), silicoctene is composed of eight-membered or four-membered rings periodically repeated along \textit{a} and \textit{b} directions, and the eight-membered rings are linked by Si-Si bonds. The Si-Si bonds can classified into two types, one type is those forming four-membered rings (\textit{b}$_{si1}$), and the other is the bonds linking the four-member rings (\textit{b}$_{si2}$). The bond lengthes of \textit{b}$_{si1}$ and \textit{b}$_{si2}$ are $\sim$2.30 and $\sim$2.25 \AA, respectively. It clearly shows that the length of \textit{b}$_{si1}$ is a bit longer than that of \textit{b}$_{si2}$. This can be explained by the fact that the overlap of Si orbitals around \textit{b}$_{si1}$ is some weaker than that around \textit{b}$_{si2}$, the heavier overlap of Si orbitals would lead to strong Coulomb attractive force, giving rise to the shortening of corresponding bonds. Additionally, the lengthes are similar with those of $\sim$2.28 \AA in silicene calculated in this work, and the slight difference is caused by the different bond angle between silicoctene and silicene. The buckling $\Delta$Z is $\sim$0.48 \AA, this value is similar with 0.45 (given by this work) or 0.44 \AA\cite{Cahangirov} of silicene.

\begin{figure}[htbp]
\centering
\includegraphics[width=8.5cm]{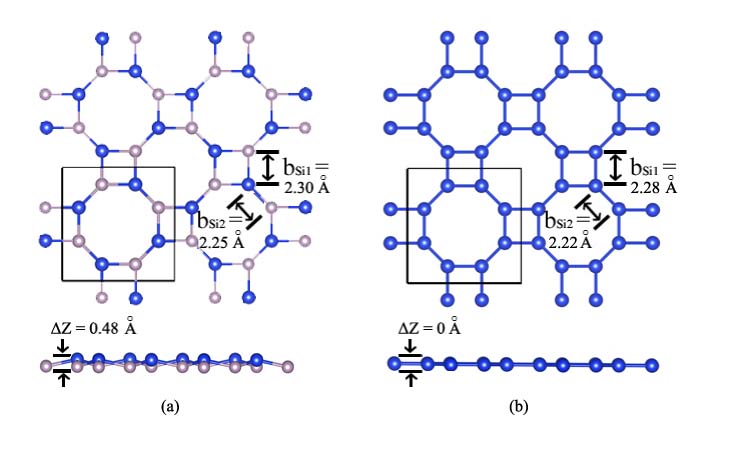}
\caption{(Color online). The top and side views of optimized structures of silicoctene. (a) is the buckling one, the brown and blue spheres represent the two silicon sublattices. (b) represents the plane one. Some parameters are listed as well.}
\label{fig:Figure1}
\end{figure}

In order to explore the possibility of experimental synthesis of silicoctene, the thermodynamic stability is discussed through the cohesive energy. The cohesive energy is expressed as the follows:

\begin{equation}
\Delta E_{silicoctene}=(nE_{Si}-E_{silicoctene})/n
\label{eq:fn}
\end{equation}

where \textit{E}$_{silicoctene}$ is the total energy of one cell of silicoctene used in this work, \textit{E}$_{Si}$ is the total energy of the isolated single Si atom, \textit{n} is the number of silicon atoms of silicoctene. The calculated result reveals that $\Delta$\textit{E}$_{silicoctene}$ is 4.58 eV/atom. This value is some smaller than 4.77 eV/atom of silicene, but larger than those of 4.42 and 4.43 eV/atom of the 2D silicon allotropes reported by Cahangirov \textit{et al}\cite{Cahangirov2} and much larger than those ($<$$\sim$4.10 eV/atom) of the ones given by Matusalem \textit{et al}\cite{Matusalem}. This result illustrates that it is possible for silicoctene to exist stably. To support this opinion, the electron localization functions (ELF) described in Fig. 4 was given to investigate the Si-Si bond. It clearly demonstrates that there exists strong interaction among the neighboring Si atoms, and the bonds present $\sigma$ state in plane and $\pi$ state out of plane.

\begin{figure}[htbp]
\centering
\includegraphics[width=8.5cm]{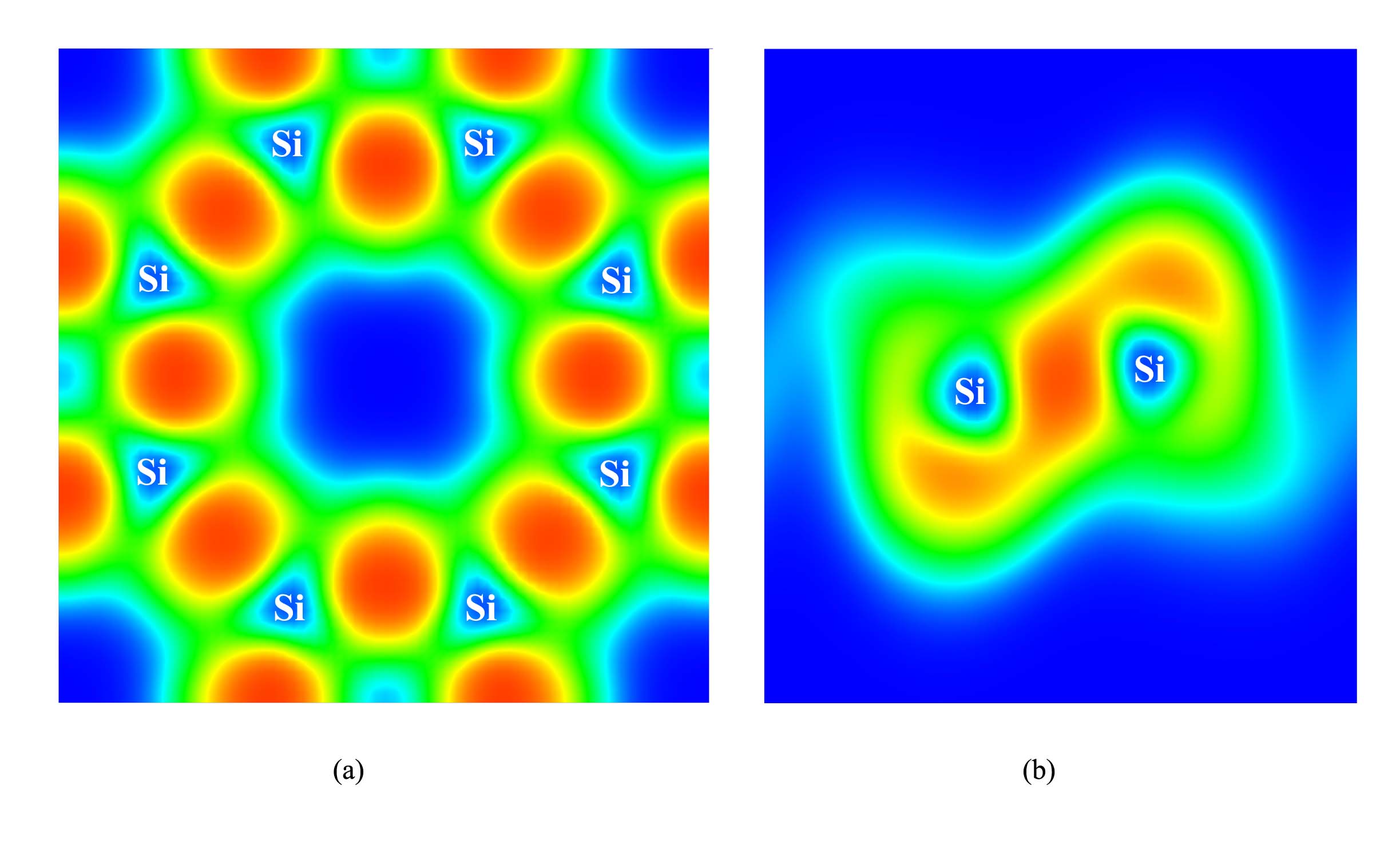}
\caption{(Color online). The calculated electron localization functions (ELF) of silicoctene. (a) and (b) are for the (001) and (100) plane, respectively.}
\label{fig:Figure2}
\end{figure}

Further, the dynamical properties of silicoctene are investigated by calculating the phonon dispersion, and Fig. 3 draws the phonon dispersion curves. If there is no imaginary frequency, it demonstrates that the structure is dynamically stable. Otherwise, the structure is dynamically unstable. Fig. 3(a) clearly shows that no imaginary frequency exists for the buckling silicoctene, which suggests the dynamical stability of this structure. Besides, the highest frequency is around 600 cm$^{-1}$ (not plotted in Fig. 3), this value is similar with that of silicene, but it is much higher than those of multi atomic layers thick 2D silicon allotropes\cite{Cahangirov2,Gimbert,ZWang}. Owing to the fact that high frequency usually means strong bonds, it demonstrates that, relative to those in multi atomic layers thick 2D silicon allotropes, there exhibit much stronger Si-Si bonds in single atomic layer thin silicoctene.

\begin{figure}[htbp]
\centering
\includegraphics[width=8.5cm]{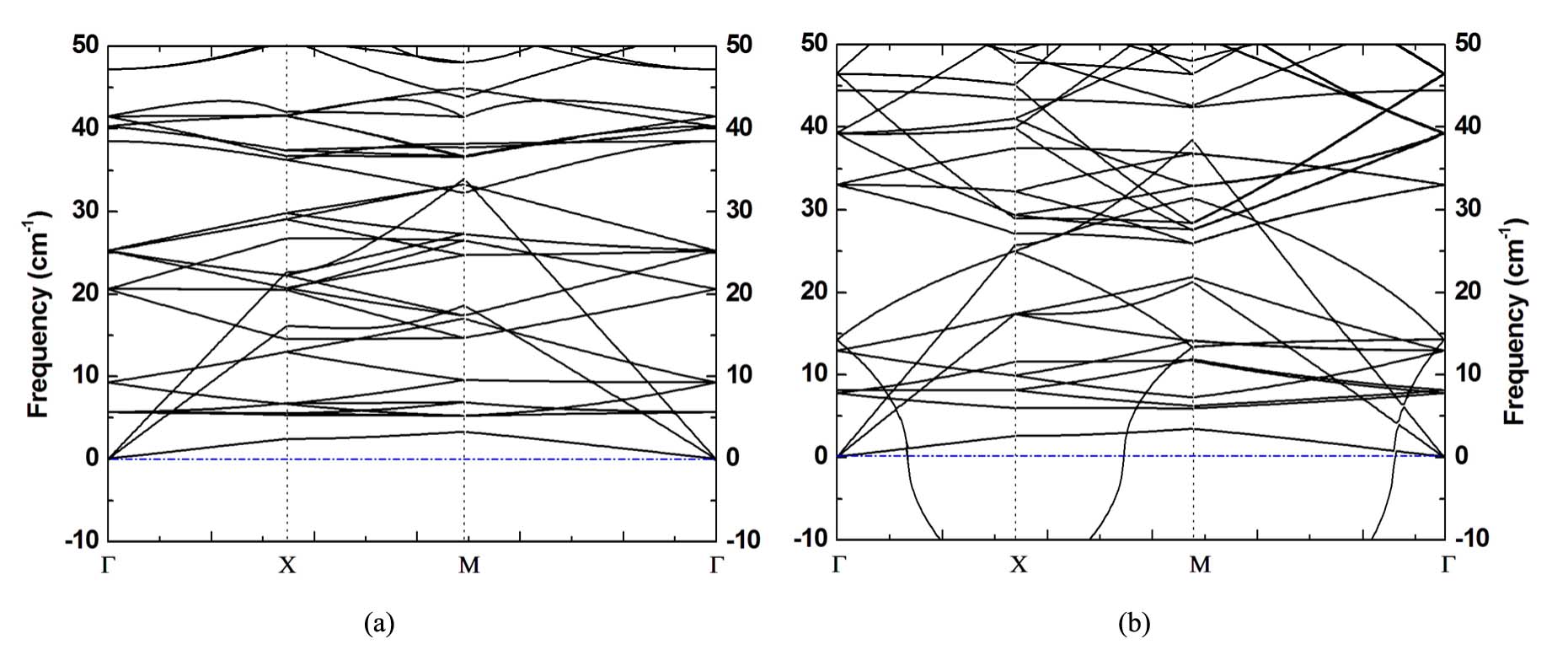}
\caption{(Color online).  The phonon dispersion curves for silicoctene. (a) and (b) are for the buckling and plane structures, respectively.}
\label{fig:Figure3}
\end{figure}

To further evaluate the thermal stability of silicoctene, the ab initio molecular dynamics (AIMD) simulations are performed. During the calculations, a large 4$\times$4$\times$1 supercell based on primitive cell is employed, AIMD simulations are calculated using the NVT ensembles, the temperature is controlled by the Nos\'{e}-Hoover method and ranged from 300 to 1100 \textit{K}, and the simulations last for 10 ps with a time step of 2.0 fs. Simulation snapshots of the last step at different temperature are described in Fig. 4. It clearly shows from the top view that there is no breaking of the bonds and the original configuration is well kept even at high temperature of 1000 K, but the deformation of rings increases with increasing the temperature. The side view shows that the warp in the atomic plane enhances with increasing the temperature. When the temperature reaches 1100 \textit{K}, the original atomic configuration of eight-membered rings is completely destroyed and the atoms are clustered together. This result illustrates that silicoctene could not exist when the temperature is above 1100 \textit{K}, and it also means that that the melting point of silicoctene is between 1000 and 1100 \textit{K}. As a result, the above performance notes that silicoctene possesses high thermal stability, suggesting its potential applications even at extremely high temperature.

\begin{figure}[htbp]
\centering
\includegraphics[width=8.5cm]{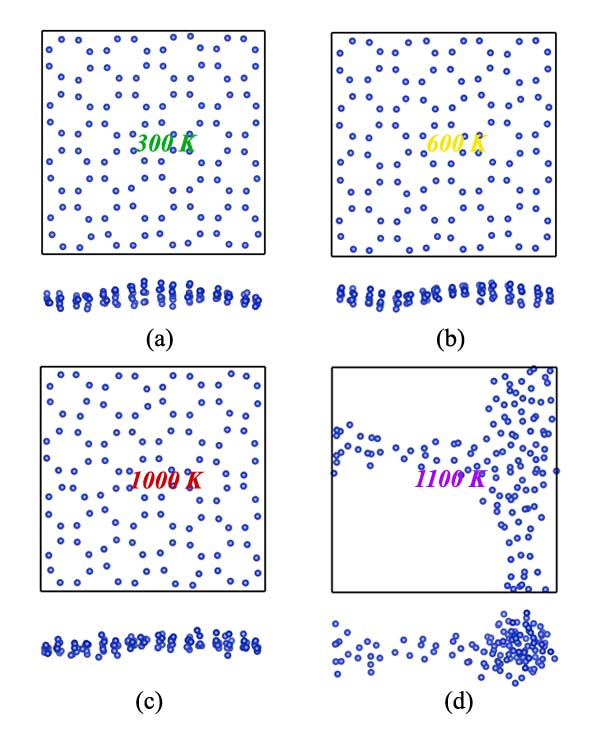}
\caption{(Color online). Snapshots of silicoctene equilibrium structures at (a) 300 \textit{K}, (b) 600 \textit{K}, (c) 1000 \textit{K}, and (d) 1100 \textit{K} at the last step of 10 ps AIMD simulations.}
\label{fig:Figure3}
\end{figure}

Due to \textit{sp}$^{3}$ hybridization of Si orbitals, we think the plane structure of silicoctene is unstable. To confirm this prophecy, some properties of the plane silicoctene was investigated as well. The optimized structure, as shown in Fig. 1(b), tells \textit{b}$_{si1}$ and \textit{b}$_{si2}$ are 2.28 and 2.22 \AA, respectively, this performance can be explained by the same rule given in the buckling silicoctene. As hypothesized, the calculated total energy is $\sim$0.03 eV/atom higher than that of the buckling one, and the calculated phonon dispersion plotted in Fig. 3(b) clearly shows that there are some imaginary frequencies with the highest value of $\sim$ -10 cm$^{-1}$. This fact confirms that the plane silicoctene is dynamically unstable.

Finally, encouraged by the stability of buckling silicoctene, the electronic structure is investigated in detail. The density of electronic states (DOS) and partial density of electronic states (PDOS) are described in Figs. 5(a). The DOS clearly shows that the silicoctene behaves as semimetallic nature with a nearly zero energy gap. The PDOS presents that, in the energy range from -1.00 to 1.00 eV, the vast majority of electronic states are composed of Si 2\textit{p$_{z}$} state. In order to deeply explore electronic structure, the band structure is pictured in Fig. 5(b). The band structure clearly shows a linear energy dispersion near \textit{E}$_{F}$ and the conduction and valence bands cross at the same reciprocal point, demonstrating that the charge carriers can transport like massless Dirac fermions. This special behavior is consistent with that of silicene reported by Cahangirov \textit{et al}\cite{Cahangirov}. To reveal the reason of this special property, the partial (band decomposed) charge density of valence and conduction band edges is plotted in Fig. 6. It shows that the charge is located on the two sides of silicoctene plane, and the valence and conduction bands exhibit obvious $\pi$ and $\pi$$^{\ast}$ bond characteristics, respectively. Associated with the characteristic of ELF and PDOS, $\pi$ and $\pi$$^{\ast}$ bands are both formed by Si \textit{p$_{z}$} orbital.

To further verify the being of Dirac fermions, the Fermi velocity in silicoctene is estimated. By neglecting the second and higher order terms with respect to \textit{q}$^{2}$, the Fermi velocity can be defined as the following equation at \textit{k} = \textit{L} + \textit{q}:
\begin{equation}
v_{F}=E(\overrightarrow{q})/\hbar|\overrightarrow{q}|
\label{eq:fn}
\end{equation}
where \textit{k} is the wave vector, \textit{L} is the location of the Dirac point, and the calculated \textit{v}$_{F}$ is $\sim$0.45$\times$10$^{6}$ ms$^{-1}$. This value is some smaller than $\sim$1.20$\times$10$^{6}$ ms$^{-1}$ in the buckled silicene generated on Ag(111) reported by Chen \textit{et al}\cite{Chen}. The large Fermi velocity further demonstrates that the charge carriers in silicoctene show the nature of massless Dirac fermions. The slight difference of Fermi velocity between silicoctene and silicene is caused by the different degree of buckling. The buckling of silicoctene is 0.48 \AA, larger than 0.45 \AA in silicene, and the transporting velocity would be decreased with increasing the buckling degree.

\begin{figure}[htbp]
\centering
\includegraphics[width=8.5cm]{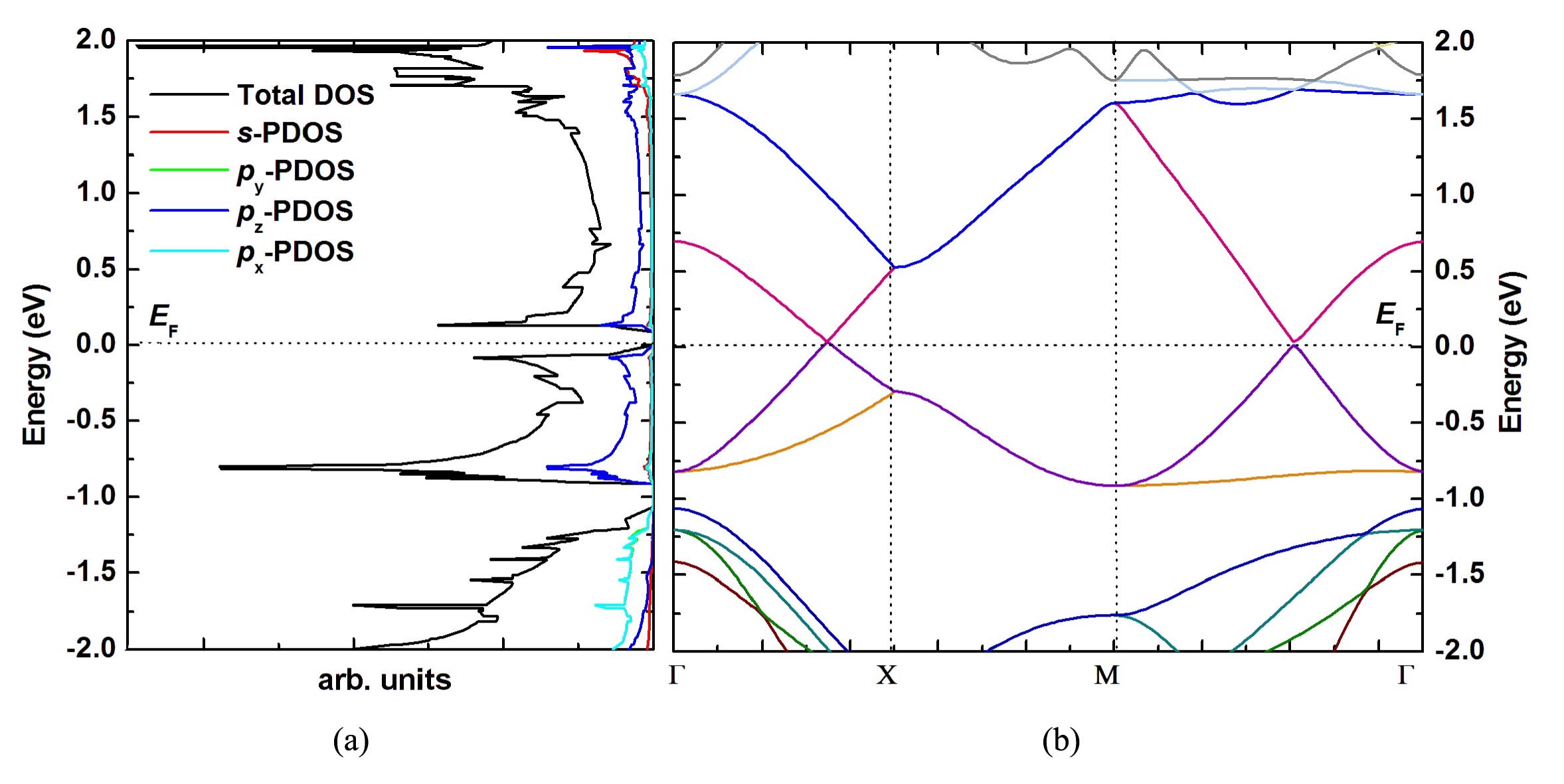}
\caption{(Color online). (a) The density of electronic states (DOS) and partial density of electronic states (PDOS) for silicoctene. (b) The band structures for silicoctene.}
\label{fig:Figure5}
\end{figure}

\begin{figure}[htbp]
\centering
\includegraphics[width=8.5cm]{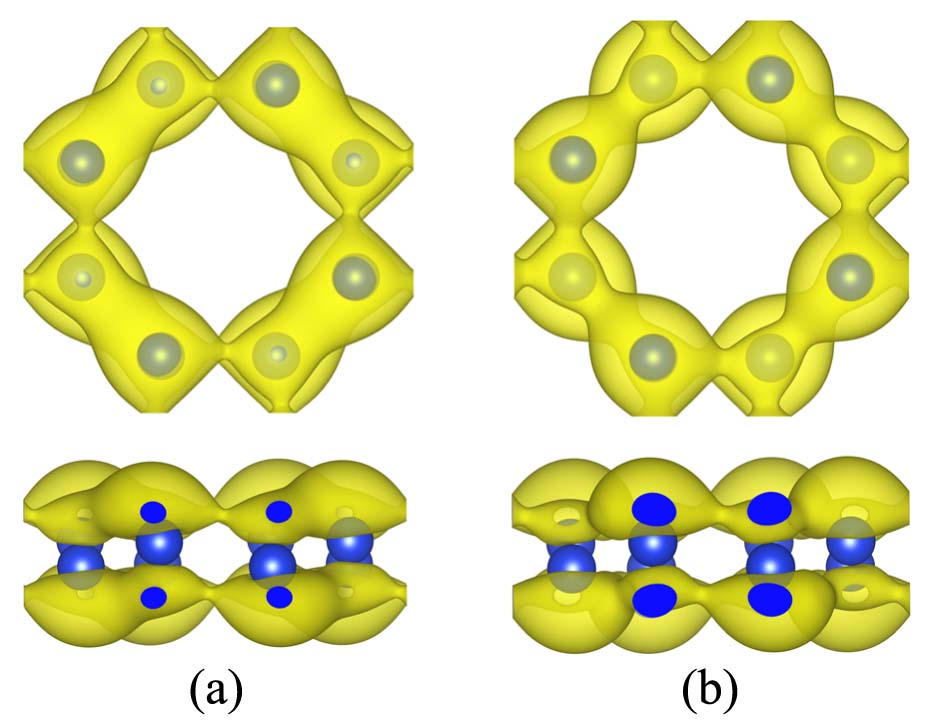}
\caption{(Color online). The partial (band decomposed) charge density for valence band (a) and conduction band (b) edges. The up and bottom ones are for top and side views, respectively.}
\label{fig:Figure6}
\end{figure}

\section{Conclusions}
Summary, via first principles calculations in combination with a swarm structure search method, silicoctene, a new single atomic layer thin silicon allotrope, is found. Silicoctene is constructed by eight-membered (or four-membered) rings with buckling nature, and the rings are linked by strong Si-Si bonds. The original configuration of silicoctene can be kept even at extremely high temperature up to 1000 K. The electronic calculation demonstrates that silicoctene behaves as semimetal with zero energy gap at \textit{E}$_{F}$. Excitedly, there presents a linear energy dispersion near \textit{E}$_{F}$, indicating that the charge carriers can transport as massless Dirac fermions confirmed by the estimation of Fermi velocity.

\vspace{1ex}
\begin{acknowledgments}
This work was supported by the National Natural Science Foundation of China (Grant nos. 11404168, 11304155, and 11374160).
\end{acknowledgments}


\begin{thebibliography}{99}
\bibitem{Novoselov} K. S. Novoselov, A. K. Geim, S. V. Morozov, D. Jiang, Y. Zhang, S. V. Dubonos, I. V. Grigorieva, and A. A. Firsov, Science {\bf 306}, 666 (2004).
\bibitem{Neto} A. H. Castro Neto, F. Guinea, N. M. R. Peres, K. S. Novoselov, and A. K. Geim, Rev. Mod. Phys. {\bf 81}, 109 (2009).
\bibitem{Cahangirov} S. Cahangirov,1 M. Topsakal,1 E. Aktu¨rk,1 H. S¸ ahin,1 and S. Ciraci, Phys. Rev. Lett. {\bf 102}, 236804 (2009)
\bibitem{Chen} L. Chen, C. C. Liu, B. J. Feng, X. Y. He, P. Cheng, Z. J. Ding,S. Meng, Y. G. Yao and K. H. Wu, Phys. Rev. Lett., {\bf 109}, 056804 (2012).
\bibitem{Lalmi} B. Lalmi, H. Oughaddou, H. Enriquez, A. Kara, S. Vizzini, B. Ealet and B. Aufray, Appl. Phys. Lett., {\bf 9}, 223109 (2010).
\bibitem{Feng} B. Feng, Z. Ding, S. Meng, Y. Yao, X. He, P. Cheng, L. Chen and K. Wu, Nano Lett., {\bf 12}, 3507 {2012}.
\bibitem{Vogt} P. Vogt, P. D. Padova, C. Quaresima, J. Avila, E. Frantzeskakis, M. C. Asensio, A. Resta, B. Ealet and G. L. Lay, Phys. Rev. Lett., {\bf 108}, 155501 (2012).
\bibitem{Fleurence} A. Fleurence, R. Friedlein, T. Ozaki, H. Kawai, Y. Wang and Y. Yamada-Takamura, Phys. Rev. Lett., {\bf 108}, 245501 (2012).
\bibitem{Meng} L. Meng, Y. L. Wang, L. Z. Zhang, S. X. Du, R. T. Wu, L. F. Li, Y. Zhang, G. Li, H. T. Zhou, W. A. Hofer and H. J. Gao, Nano Lett., {\bf 13}, 685 (2013).
\bibitem{Huang} L. Huang, Y. F. Zhang, Y. Y. Zhang, W. Y. Xu, Y. D. Que, E. Li, J. B. Pan,Y. L. Wang, Y. Q. Liu, S. X. Du, S. T. Pantelides, and H. J. Gao, Nano Lett. {\bf 17}, 1161 (2017).
\bibitem{Gimbert} F. Gimbert, C. Lee, R. Friedlein, A. Fleurence, Y. Yamada-Takamura, and T. Ozaki, Phys. Rev. B {\bf 90}, 165423 (2014).
\bibitem{Cahangirov2} S. Cahangirov, V. Ongun \"{O¨ }z\c{c}lik, A. Rubio, and S. Ciraci, Phys. Rev. B {\bf 90}, 085426 (2014).
\bibitem{Matusalem} F. Matusalem, M. Marques, L. K. Teles, and F. Bechstedt, Phys. Rev. B {\bf 92}, 045436 (2015).
\bibitem{ZWang} Z. H. Wang, M. W. Zhao, X. F. Zhou, Q. Zhu, X. M. Zhang, H. F. Dong, A. R. Oganov, S. M. He, and P. Gr\"{u}nberg, http://arxiv.org/abs/1511.08848 (2015).
\bibitem{Wu} H. P. Wu, Y. Qian, S. H. Lu, E. J. Kan, R. F. Lu, K. M. Deng, H. Wang, and Y. M. Ma, Phys. Chem. Chem. Phys., {\bf 17}, 15694 (2015).
\bibitem{Ivanovskii} A. L. Ivanovskii, Russian Chemical Reviews {\bf 81}, 571 (2012).
\bibitem{wang1} Y. Wang, J. Lv, L. Zhu, and Y.M. Ma, Phys. Rev. B {\bf 82}, 094116 (2010).
\bibitem{wang2} Y. Wang, J. Lv, L. Zhu, and Y.M. Ma, Comput. Phys. Commun. {\bf 183}, 2063 (2012).
\bibitem{Perdew} J. P. Perdew, K. Burke, and M. Ernzerhof, Phys. Rev. Lett. {\bf 77}, 3865 (1996).
\bibitem{Kresse1} G. Kresse and D. Joubert, Phys. Rev. B: Condens. Matter Mater. Phys. {\bf 59}, 1758 (1999).
\bibitem{Kresse2} G. Kresse, and J. Furthm\"{u}ller, Comput. Mater. Sci. {\bf 6}, 15 (1996).
\bibitem{K} K. Parlinski, Z. Q. Li, Y.Kawazoe, Phys.Rev.Lett.{\bf 78}, 4063 (1997).
\bibitem{A} A. Togo, F. Oba, I. Tanaka, Phys.Rev.B {\bf 78}, 134106 (2008).
\end{thebibliography}
\end{document}